# Cyclic Strength and Nonlinear Material Fracture Mechanics
# (by the example of steels)

L.S.Kremnev

*Moscow State Technological University «Stankin», Moscow, Russia*
kremnevls@yandex.ru

**Abstract** – It has been shown that a material fatigue fracture diagram can be viewed as a locus of points with their $\sigma$ and $\sqrt{l}$ coordinates' product equaling to $K_{1c}/2$ and their $\sigma$ and $l$ product – to $G_{1c}/2$, where $K_{1c}$ and $G_{1c}$ are respectively the force and energy criteria of non-linear fracture mechanics. It has been established that the average number of interatomic bonds destroyed within a single alternate stress $\nabla_{1cs}$ cycle is directly proportional to $\sigma$ which is twice as large as a peak value of $\sigma^a$. It has been established that the low-cycle fatigue is characterized by the fact that $\sigma > \sigma_{0.2}$ and $\nabla_{1cs} > 1$, high-cycle fatigue - by $\sigma = \sigma_{0.2}$ and $\nabla_{1cs} = 1$, and giga-cycle fatigue – by $\sigma < \sigma_{0.2}$ and $\nabla_{1cs} < 1$. An individual interatomic bond cannot be destroyed part by part but as a single unit. The latter circumstance means that in case of giga-cycle fatigue a single interatomic bond is destroyed within several (a number) of cycles rather than within a single cycle. The factors F - collapsibility and R - resistibility have been proposed and referred to as essential material physical constants. The introduced notion $\nabla_{1cs}$ and the established linear nature of $\nabla_{1cs}$ – relationship make it possible to: a) clarify the physical nature of the fatigue crack growth in low-, high- and giga-cyclic fracture zones; b) determine the nature of a disruption appearing on a fatigue fracture diagram; c) plot the fatigue fracture diagram on the basis of the results obtained during a single specimen cyclic strength testing with a selected value of $\sigma \geq \sigma_{0.2}$. In case of giga-cycle fatigue it is important (with the same purpose in mind) to determine this dependence for $\sigma < \sigma_{0.2}$. It is recommended to use the criterion $G_{1c}$ for calculating the value of length $l_{cr}$ which in contrast to $K_{1c}$ has a clear physical nature.

*Key words: cyclic strength, fracture mechanics criteria, fatigue tests.*

Determination of product service life period preceding its destruction depending on the value of the cyclically varying stress applied is often referred to as one of the main problems to be solved by designers. In this connection it is important to know the material fatigue strength limits, i.e. its peak stress values $\sigma^a$, at which the material is not destroyed for a predetermined period of product service life. For example, in case of iron-based alloys (steel, cast iron) this value comprises $10^7$ cycles ; for copper-based alloys (brass, bronze) – $10^8$ cycles . Under alternate stress influence the fatigue crack grows up to its critical length $l_{cr}$. relatively slowly, after which it quickly propagates throughout the product , thus forming a rupture area. The described fatigue fracture mechanism is well known, since in practice it takes place more often than the mechanism of destruction caused by influence of a permanently applied load. The material fracture mechanisms observed under variable or constant loading are different, nevertheless they have some common features. The greatest difference between them lies in the fact that under alternate load conditions the material disrupts for a finite number of cycles under the stresses that are considerably smaller than its tensile strength (low cycle fatigue) and even smaller than the yield stress. If the stress is less than the yield strength (elasticity) limit, than the fracture behavior changes qualitatively. Fracture appears under stresses with the number of cycles several orders of magnitude greater than in process of low-cycle fatigue (high-cycle fatigue phenomenon) [1]. The difference between low-and high- cycle fatigue can be explained by the fact that in the first case the crack grows rapidly under the influence of plastic deformation. In the second case plastic deformation is absent or moderate thus restricting the crack growth. This fact may serve as an explanation of a considerably smaller crack growth (except for its critical length) at a constant (rather than at variable) loading, in the half-cycle compression of which the plastic shear deformation under the influence of tangential stress stimulates the fatigue cracks growth. However, if we turn to the typical fatigue fracture diagrams, e.g. those of steels influenced by the most typical symmetric loading cycle, it becomes possible to observe some common details of material destruction taking place under constant and variable loading conditions.

Experimental diagrams of steel fatigue fracture (Fig. 1) are well approximated by equations (1) and (2) respectively:

$$2\sigma^a = A / l_{cr}, \qquad (1)$$

$$2\sigma^a = B / \sqrt{l_{cr}}, \qquad (2)$$

where $\sigma^a$ - peak cyclic stress value, A and B – factors (material constants).

In [2] (see also Application) it has been shown that the force and energy criteria of nonlinear mechanics of elastic-plastic material fracture (NMMF) observed under static loading are referred to as crack growth resistance $K_{1c}$ and fracture toughness $G_{1c}$:

$$K_{1c} = 2\sigma \sqrt{l_{cr}} \qquad (3)$$

$$G_{1c} = 2\sigma \cdot l_{cr} = J_{1c}. \qquad (4)$$

Here $J_{1c}$ is Rice integral [7].

Due to the fact that $\sigma$ is equal to maximal tensile stress affecting the material during the of alternate stress ($\sigma$) half-cycle, then

$$\sigma = 2\sigma^a. \qquad (5)$$

From eq. (1) and (2) it follows that

$$B = K_{1c} / 2, \qquad (6)$$

$$A = G_{1c} / 2 \qquad (7)$$

Consequently, the material fatigue fracture diagrams in coordinates " $\sigma - \sqrt{l}$", "$\sigma - l$", (Fig. 1) are the locus of points with the coordinates providing the possibility to determine the values of stress $\sigma$ and crack length $l_{cr}$ responsible for its destruction.

This finding is consistent with the physical nature of the problem discussed, according to which the criteria for NMFM destruction $K_{1c} = 2\sigma \sqrt{l_{cr}}$ and $G_{1c} = 2\sigma l_{cr}$ can be treated material constants, with their values being independent of each factor taken separately, but determined by their product.

Cracks (the computer model of one of them is presented in Figure 2 [4]) are located in the most densely-packed crystal lattice planes located far apart from each other . Therefore, the atomic bond energy between these planes is smaller than that between the planes with lesser density of atomic packing.

Figure 3 [5] shows a fatigue crack in the material under cyclic stress conditions with the peak (amplitude) value F(σ). Interatomic bond stiffness n at the crack tip is smaller, and its elongation – larger than those of bond n +1. It cn be easily seen that there are no bonds to the left of n bond, the presence of which (prior to their destruction) could increase its resistance to stress $2\sigma^a$. Thus, the bond n is destroyed prior to bond n +1, the latter being sure to retain with a $2\sigma^a$ decreasing in a subsequent half-cycle i.e. in the course of compression. It transits into the position of n bond observed prior to its destruction. Thus, perhaps, during each of the stress cycles the rupture of one interatomic bond and, hence, a crack buildup for one respective crystal lattice spacing takes place. This speculative conclusion can be proved by calculations. Firstly, it should be noted that a designer who is to estimate a product strength shall proceed from the fact that the maximum permissible acting static stress $\sigma^{max}$ in its most dangerous section can not be greater than the ultimate yield stress $\sigma_{0.2}$. To be more precise, $\sigma^{max} = \sigma_{0.2}/ n$, where n is a safety factor (n>1).

Let us turn to a group of heat hardenable structural steels commonly used in mechanical engineering, with their carbon concentration varying within 0.30 - 0.50%, and the alloying components percentage varying within a wider range. Let's select a conventional steel grade 40X ($\sigma_B$ = 1000Mpa, $\sigma_{0.2}$= 800Mpa, KCU = 0.6 MJ/m$^2$, $K_{1c}$=73MPam$^{1/2}$) [2 , 6].

Consider the toughest scheme of alternate symmetric cyclic loading: n = 1. From equation (3) we find:

$$l_{cr} = 0.25\ ( K_{1c}/\sigma_{0.2})^2 \tag{8}$$

Steel 40X , according to (8) has $l_{cr}$ =2.08·10$^{-3}$ m. The average value of the destroyed atomic bond Δ between the most densely-packed lattice planes observed during 1 alternate stress cycle can be found from next formula

$$\Delta = l_{cr} / N, \tag{9}$$

where N = 10$^7$ cycles accepted (basic) prior to failure. It can be seen that the value of the bond destroyed Δ = 2.08·10$^{-10}$ m. The distance between the planes {110} with maximal atomic density in the Fe$_\alpha$ crystal lattice equals to 2.03·10$^{-10}$ m [6]. One can see that the length of a crack located in {110}) iron lattice plane increases by the same value. Such a coexistence of the results obtained could be considered occasional taking into consideration the experimental errors in determining values of $K_{1c}$ and $\sigma_{0.2}$ assumed for the calculations. Nevertheless, it does not contradict with the conclusion that for every cycle tensile strength $2\sigma^a \approx \sigma_{0.2}$ the crack length increases by one interplane spacing Δ. Along with these, it is obvious that with $2\sigma^a > \sigma_{0.2}$ not one, but many bonds are destroyed, and the more of them are destroyed the higher is the peak stress $\sigma^a$ value.

The number of atomic bonds [b] destroyed during a single cycle [cl] of loading, $\nabla_{1cs}$ [b/cl], for the total number of which (on the average) the crack respectively enlarges in Fe$_\alpha$ – base alloys can be determined in the following way:

$$\nabla_{1cs} = l_{cr}/\ 2.08 \cdot 10^{-10}\,N, \tag{10}$$

where $l_{cr}$ can be found from (8) and N, the up-to-failure number of cycles up to the specimen failure under the selected stress $\sigma^a$ action can be determined experimentally.

It should be remembered that if $2\sigma^a$ the ultimate stress is observed during a half-tensile cycle.

Equation (10) establishes a relationship between the experimental number of prior-to-fracture cycles N and the calculated crack length $l_{cr}$, i.e. between the fatigue diagrams (Fig. 1a and 1c). In the Table below one can see the fatigue tests results ($\sigma^a$ and N) relating to a conventional steel grade 40HNMA. The Table also contains the calculation results concerning $\nabla_{1cs}$ equations (8) and (10). Figure 5 demonstrates graphic representation of $\sigma^a$ and $\nabla_{1cs}$ functional relationship.

Table

**The number of broken interatomic bonds $\nabla_{1cs}$ during 1 cycle of alternate peak stress $\sigma^a$ depending on its value**

| № | $\nabla_{1cs}$, BL/C | $\sigma^a$, MPa | $l_{cr}$, m | N number of cycles up to fracture |
|---|---|---|---|---|
| 1 | 97 | 628 | $12.2\times10^{-4}$ | $6\times10^4$ |
| 2 | 68 | 590 | $13.9\times10^{-4}$ | $1\times10^5$ |
| 3 | 42 | 530 | $17.2\times10^{-4}$ | $2\times10^5$ |
| 4 | 30 | 509 | $18.7\times10^{-4}$ | $3\times10^5$ |
| 5 | 20 | 480 | $20.0\times10^{-4}$ | $5\times10^5$ |
| 6 | 15 | 470 | $21.9\times10^{-4}$ | $7\times10^5$ |
| 7 | 11 | 462 | $22.7\times10^{-4}$ | $1\times10^6$ |
| 8 | 1 | 445 | $24.4\times10^{-4}$ | $1\times10^7$ |

Steel 40HNMA after heat-hardening, $\sigma_B = 1030$ MPa; $\sigma_{0.2} = 885$ MPa; $K_{1c} = 88$ MPa·m$^{0.5}$ [2, 5]. Figure 4 shows that the characteristic of a crack growth average rate $\nabla_{1cs}$ is directly proportional to the peak stress $\sigma = 2\sigma^a$. The result obtained is natural and quite predictable.
This dependence proves the fact that the coefficient in equation (П2) of the Appendix is actually equal to ½.

From Figure 4 it follows:

$$\nabla_{1csi} \approx \nabla_{1cs1} (\sigma^a_i - 0.5\ \sigma_{0.2})/(\sigma^a_1 - 0.5\ \sigma_{0.2}) \qquad (11)$$

where $\sigma^a_1$ and $\sigma^a_i$ are the peak stresses of a real and hypothetic specimens, respectively. This relationship allows us to find any number of points and use them for drawing up a material fatigue fracture diagram (Fig. 1a) on the basis of a single experiment results. It determines the number of cycles $N_1$ up to a standard specimen fracture under the influence of an alternating stress with a selected amplitude value $\sigma^a_1$.
For doing these using the value $l_{cr}$ calculated by equation (8) and experimentally obtained value of $N_1$ one can find the value $\nabla_{1cs1}$ using equation (10). Further, in accordance with equation (11), one can calculate the value of $\nabla_{1csi}$ (i = 2, 3, 4, ..) . On the basis of equation (10), knowing the values of $l_{cr}$ (see equation (8) and $\nabla_{1csi}$ it is possible to obtain the number of cycles $N_i$ preceding the fracture of hypothetical (i- the specimen under the action of a selected stress $\sigma^a_i$.). Using the selected values of $\sigma^a_i$ and calculated values of $N_i$ one can obtain the points necessary for charting a material fatigue fracture diagrams within "$\sigma^a_i$ - $N_i$" coordinates respectively (Fig.1b) . If one draws a sloping straight line by two points ($\nabla_{1cs1}$ and a point with coordinates $\nabla_{1cs} = 1$ and $\sigma = \sigma_{0.2}$ on the abscess of the diagram shown in Fig.4) it will be particularly easy to find the values of $\nabla_{1csi}$.

For brittle materials with $\sigma_{0.2} \approx \sigma_B$, the alternative version consists in drawing an inclined line in accordance with the fatigue test results of two specimens at $\sigma_1 > \sigma_2 > \sigma_{0.2}$. In selecting the points' coordinates and for reducing the test duration it is expedient to test the specimens at comparatively large $\sigma^a$ (that is for not very large number of cycles up to fracturing). If the linear dependence between $\nabla_{1cs1}$ and $\sigma$ also remains valid at $\sigma < \sigma_{0.2}$, than the possibility may appear to determine a material giga-cycle fatigue strength on the basis of a single specimen low-cycle fatigue test results. Therefore it is essential to determine the nature of this dependence in the giga-cycle fatigue zone. There are grounds for believing that the linear dependence considered remains valid in the giga-cycle fatigue zone as well. Nevertheless these hypothetical conclusions require experimental confirmation

The results of fatigue tests are characterized by considerable scattering. Therefore, to obtain statistically reliable data it is necessary to destroy up to several tens of standard specimens in order to obtain a single point of a fatigue diagram. There is a possibility that in the case considered to achieve this goal it will be necessary to obtain a statistically reliable ( with a probability of , for example, 95% or 99%) position of one point in order to adequately calculate and built a certain material fatigue fracture diagram as a whole using the scheme proposed. It would be advisable to experimentally confirm the implementation of this possibility. The second feature, demonstrated in Figure 4 is in the fact that the endurance limit of steel 40HNMA as well as of that of steel 40X, is equal to its yield point $\sigma_{0.2}$. This fact confirms the opinion of a number of authors [1] who claim that such conformity is characteristic of many materials. By the way, many nonferrous alloys have $\sigma_{0,2}$ noticeably smaller than that of ferrous alloys. That's why the acceptable crack length $l_{cr}$ of the latter is often larger than that of the former in accordance with equation (8) , and , consequently, their base number of up-to-failure cycles N shall be increased from $10^7$ to, for example, $10^8$.

It should be considered that the factors of collapsibility $F = \nabla_{1cs} / \sigma$ and the fracture resistance power $R = \sigma / \nabla_{1cs}$ are important loaded material constants

One should also bear in mind that if it is possible to build diagrams of materials fatigue fracture with the help of known values[1] of $K_{1c}$, then, consequently, these diagrams, the larger part of which have been obtained experimentally and widely presented in the literary sources, can be used to to determine the criteria of material fracture mechanics without performing relevant experiments. As it turned out ,the fact that the diagrams of material fatigue fracturing and NLFM criteria are closely interconnected (Fig. 1), provides the opportunity to judge on the reliability of experimental data ($\sigma^a$, N and $l_{cr}$), on the basis of which the diagrams were built. Indeed, it should be expected that the products of coordinates $\sigma^a$ and $l_{cr}$ (see Fig.1, Table) of all the diagram points shall be similar. Statistical processing of the data presented in the Table has shown that the average value of $\sigma^a \cdot l = 937 \pm 80$ MPam with a probability of 95 % (minimum and maximum values of the products were not considered – Table position No1 and No 8).

The considered problem of identifying the material fatigue failure diagram coordinates goes beyond their frames because it is connected with the criteria of material fracture mechanics as a whole. One should remember that the force and energy criteria of fracture mechanics $K_{1c} = 2\sigma \cdot \sqrt{l}$ and $G_{1s} = (K_{1c})^2/E$ were obtained in their conventional form for the conditions of elastic( (xbrittle ) material fracture. Nevertheless, the fracture toughness $K_{1c}$ is successfully used to describe elastic-plastic fracture thanks to (as it is considered) D. Dugdale's hypothesis. In accordance with this hypothesis a thin layer of plastically deformed ( hardened) material is formed at the crack tip prior to the process of fracturing. Alongside with it, fracture toughness $G_{1c}$ isn't practically used in NLFM except for rare cases.

---

[1] Steel fracture toughness can be calculated by the following formula:
$K_{1c} = 2.38\sigma_B/\sigma_{0.2} \cdot (KCU \cdot \sigma_B)^{0.5}$ MPam$^{0.5}$ [2].

In [2] (see also the Appendix) on the basis of the obtained J-integral $J_{1c}$ J. Rice solutions [7] it was shown that the criteria of NLFM are $K_{1c} = 2\sigma \cdot \sqrt{l}$ and $G_{1c}(J_{1c}) = 2\sigma \cdot l$. Using equation as (10) one can obtain:

$$2\sigma \cdot l = 2\sigma \cdot \sqrt{l} \cdot \sqrt{l} = (K_{1c})^2/2\sigma \qquad (12)$$

Crack resistance $G_{1c} = (K_{1c})^2 / E$ represents the energy criterion of LMP and is 1-2 orders of magnitude less than the energy criterion NLFM $G_{1c}(J_{1c})$. The value $G_{1c}$ is an important physical constant of a structural material. One should bear in mind that the fracture toughness $G_{1c}$ [J/m$^2$] is the energy required to form a unit area of a growing crack.

Thus, the physical nature of the diagram branch rupture (Fig.1b), i.e. appearance of a horizontal segment on it consists in the following: above the segment, that is at $\sigma > \sigma_{0.2}$, the material exists in the cold-working strengthened condition and to destroy this material it is necessary to apply an increased $\sigma$ value. Due to the fact that for any point of a fatigue fracture diagram the value of $G_{1c} = 2\sigma \cdot l$ is constant the critical length of $l_{cr}$ crack appears to be small. The results of the circumstance considered are in the increased values of $\nabla_{1cs}$ and a high crack propagation rate. When $\sigma < \sigma_{0.2}$ that is when a material hasn't been strengthened by cold working the values of $\sigma, \nabla_{1cs}$ and a crack growth value appear to be rather small and $l_{cr}$ – large. It is worth mentioning that in the case of long-term operation of a structure material, when $\sigma < \sigma_{0.2}$ as a result of the latter reducing under the influence of Baushinger effect $\sigma > \sigma_{0.2}$ It can cause premature destruction of the structure. Unfortunately, the effect of $\sigma^a$ and N on the value of Baushinger effect has not been sufficiently studied yet.

It is obvious that the results of fatigue fracture diagram coordinates' identification (Fig.1b) obtained in this work represent the experimental evidence of the fact that the crack toughness $G_{1c} = J_{1c} = 2\sigma \cdot l = (K_{1c}/2\sigma)^2$ is the energy criterion of NLFM.

In Fig.1 it can be seen that the higher and the farther from the starting points of «$\sigma$» and «l» coordinates the fatigue diagram is the larger is the value of material fatigue fracture resistance. Therefore the value of fracture toughness $G_{1c} = 2\sigma \cdot l$ is the selection criterion of a material possessing maximal cyclic stress resistance.

Figuring out the reasons for the sharp change in the nature of the material fracture during its transition from low-cycle to high-cycle area naturally arouses the problem of the physical nature of crack propagation in the range gigacyclic fracture range.

As it was shown, at $2\sigma^a \approx \sigma_{0.2}$ one interatomic bond ($\nabla_{1cs} = 1$) disrupts per 1 cycle (on the average) which corresponds to the plastic deformation beginning because dislocation movement is caused by destruction of at least one interatomic bond. Herewith, for the crack length of to reach a critical value of $l_{cr}$ a large number of stress cycles N shall be observed up to material fracture (up to $10^7$ for iron-based alloys). Therefore, a sufficiently large horizontal section (Fig.1b) or a transition bending point appears on the chart.

If $2\sigma^a = \sigma_{el}$, then $\nabla_{1cs} < 1$, i.e. $\sigma_{el} < \sigma_{0.2}$. Since a unit interatomic bonding cannot be destroyed part by part, but as a whole, hence in this case the destruction proceeds in several cycles, not in a single one. It is obvious that the change of fracture physical mechanism leads to an abrupt increase in the allowable basic) number of cycles up to fracture. A second inclined branch appears on the diagram that propagates into the gigacyclic fatigue zone (N = $10^8 - 10^{10}$). Perhaps, in this case the physical nature of a crack growth changes from the dislocation to the vocation one.

It is obvious that the crack growth physical mechanism changes from the dislocation to the vocation one. In this case the crack growth is connected with accumulation of necessary amount of vocations at its tip. It is natural, that during this period of vocations' accumulation the crack doesn't grow.

The phenomenon of cyclic fatigue is widely known and constitutes a subject of numerous investigations. In this regard, it seems appropriate to suggest that some concepts and definitions should be clarified taking into consideration the fact that a material fatigue fracture diagram (Fig.1c) consists of three zones differing in the nature of interatomic bonds destruction taking place within their boundaries:

I – low-cycle fatigue zone, in which the maximum cycle tensile stress $\sigma > \sigma_{0.2}$. It results in deformational stress (cold work hardening) and causes a single-cycle fracturing of a number (several) interatomic bonds at the peak of a crack. These facilitate its comparatively rapid growth up to the critical length that is up to the material fracturing (within ~$10^2 - 10^6$ cycles) – dislocation crack growth zone.

II – high-cycle fatigue zone where $\sigma = \sigma_{0.2}$ and therefore causes a single interatomic bond destruction at the crack tip within 1 cycle on the average which results in its relatively slow growth up to fracturing (in ~ $10^7$ cycles for steels). It seems likely that in this case the crack grows by both dislocation and vacancy mechanism.

III – giga-cycle fatigue zone where $\sigma < \sigma_{0.2}$. This circumstance keeps the value of $\nabla_{1cs} < 1$ and therefore not one but several (set of) stress cycles are required to destroy a single interatomic bond at the crack tip. These provide a particularly slow crack growth up to its critical length, i.e. up to fracture. (in ~$10^8 - 10^{10}$ cycles or more). The crack growth proceeds by the vacancy.

Conclusions:

1. The paper determines:
   - the physical nature of a crack growth depending on the values of the cyclic stress $\sigma$ and yield limit $\sigma_{0.2}$;
   - the nature of a disruption appearing on a fatigue fracture diagram.
2. It has been experimentally proven that the fracture toughness $G_{1c} = J_{1c} = 2\sigma \cdot l = (K_{1c})^2/2\sigma$ can be referred to as the energy criterion of NLFM. The equation $G_{1c} = (K_{1c})^2/E$ is the energy criterion of LFM.
3. The values of the fracture criteria $G_{1c} = 2\sigma \cdot l$ and $K_{1c} = 2\sigma \cdot \sqrt{l}$ can be obtained on the basis of a single standard experiment results. Since $G_{1c}$ has a clear physical sense in contrast to $K_{1c}$ it is expedient to use the value $l_{cr} G_{1c}$ instead of $K_{1c}$.
4. It has been shown that the diagrams of material fatigue fracture are the locus of points with the products of their coordinates determining the force and energy criteria of nonlinear fracture mechanics : $K_{1c} = 2\sigma \cdot \sqrt{l}$ and $G_{1c}(J_{1c}) = 2\sigma \cdot l$ respectively.
5. Fracture toughness $G_{1c} = 2\sigma \cdot l$ can be referred as the selection criterion of a material with maximal fatigue strength.

6. The characteristic of crack propagation rate $\nabla_{1cs}$ [b/cl] has been obtained. It is equal to the average number of destroyed interatomic bonds of a material crystal lattice, and the total length of these bonds is the value for which the crack length correspondingly increases per one cycle of alternate peak stress $\sigma^a$.

7. The linear dependence of $\nabla_{1cs}$ on $\sigma^a$ has been established (at $\sigma \geq \sigma_{0.2}$).

8. It has been shown that the abovementioned relationship makes it is possible to build a fatigue diagram in coordinates «$\sigma^a$,N» by an arbitrarily large number of its constituent points on the basis of the results of a single fatigue fracture experiment using a standard specimen (at $\sigma \geq \sigma_{0.2}$).

9. If the linear dependence between $\nabla_{1cs1}$ and $\sigma$ at $\sigma < \sigma_{0.2}$ remains valid, the possibility appears to determine the giga-cycle fatigue strength of a material on the basis of the results of a single specimen for fatigue testing. Therefore it is quite important to determine the nature of this dependence in the giga-cycle zone.

10. The possibility of statistical evaluation of experimental data reliability has been established which is necessary for fatigue fracture diagram construction.

11. It is expedient to broaden the sphere of research into the influence of cyclic loading duration on the yield limit. Its results are sure to increase the reliability of material cyclic strength forecasts.

Acknowledgement

The author thanks prof. V. Matyunin who kindly provided him with necessary materials on steel 40HNMA.

Appendix I

Let us consider the plate with thickness δ made of an elastic-plastic material, which has been loaded with tensile stress σ (Fig.P1).Let us assume that in the plate a side crack has appeared with the length *l* and opening *a*. The destroying stress in the crack tip is referred to as a direct reason for its formation ($\sigma_y - \sigma$), whereas the crack is indirectly formed under the action of the stress σ ($\sigma_y > \sigma$ due to the concentrating influence of the crack tip). According to the energy conservation law, for the system equilibrium condition the values of destruction energies $W_y$ and W of forces $P_y$ and P, responsible for the appearance of stresses ($\sigma_y - \sigma$) and σ, are equal. Let us determine the values $W_y$, W and make them equal

$$W_y = P_y \cdot S_y, \tag{P.1}$$

where $S_y$ is the motion of force $P_y$ in the direction of axis «Y».

From Fig.**P2** it follows that

$$P_y = \tfrac{1}{2} (\sigma_y - \sigma) \cdot \delta \cdot \Delta, \tag{P.2}$$

where ($\sigma_y - \sigma$) is the fracture stress. Genuinely, at $\sigma = \sigma_y$ it is equal to 0, and the plate remains non-destroyed, but influenced by the stress σ. Due to the increase of σ and the crack growth the fracture stress value increases from 0 to ($\sigma_y - \sigma$). Hence, the average value of the stress, destroying the plate, is equal to $0.5(\sigma_y - \sigma)$. This circumstance has been taken into account in equation (P.2) by means of the factor ½. Δ - the minimal distance for which the crack edges inconvertibly diverge at its tip (Fig.P2). Thus

$$W_y = \tfrac{1}{2} (\sigma_y - \sigma) \cdot \delta \cdot \Delta \cdot a, \tag{П.3}$$

where $a = \Sigma \Delta = S_y$ (critical crack opening «CCO» in the moment of fracture).

$$W = P \cdot \Delta = \sigma \cdot l \cdot \delta \cdot \Delta. \tag{П.4}$$

By equaling equations (P.3) and (P.4), we can obtain:

$$(\sigma_y - \sigma) \cdot a = 2\sigma \cdot l \tag{P.5}.$$

The left-hand part of equation (P.5) is expression for $J_{1c}$ – Rice integral [7], and the right-hand part of it is its second expression:

$$J_{1c} = G_{1c} = 2\sigma \cdot l. \tag{P.6}$$

It should be emphasized that from (P.5) one can obtain:

$$\sigma_y = \sigma \cdot (1 + 2l/a) \tag{P.7}$$

and

$$\sigma_y = \sigma \cdot (1 + 2\sqrt{l/\rho}) \tag{P.8}$$

for a crack with an oval tip and radius $\rho$. Equations (P.7) and (P.8) are long and well known in the material fracture mechanics and have completely proven their correctness in practice.

It should be pointed out that the equation $G_{1c} = 2\sigma \cdot l$ can be calculated using the results of a standard experiment aimed at determining $K_{1c} = 2\sigma \cdot \sqrt{l}$, whereas the experimental determination of $G_{1c} = (\sigma_y - \sigma) \cdot a$ seems to be impossible.

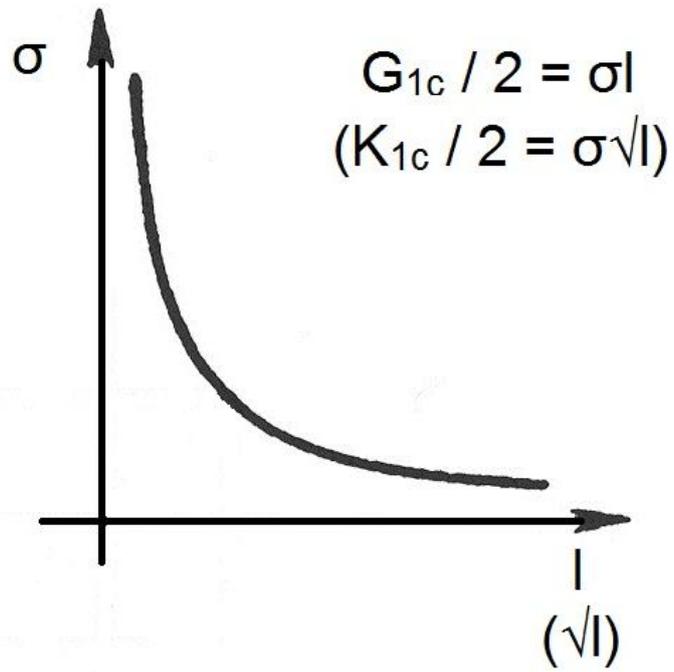

a)

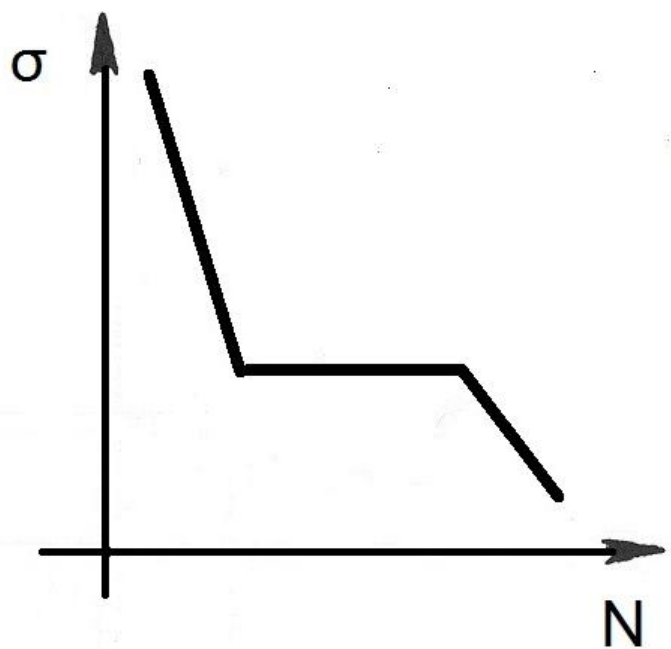

b)

Fig. 1. Typical diagrams of fatigue fracture of steel, constructed in the coordinates :
a) «$\sigma^a$, l», «$\sigma^a$, $\sqrt{l}$»,  в) «$\sigma^a$, N».

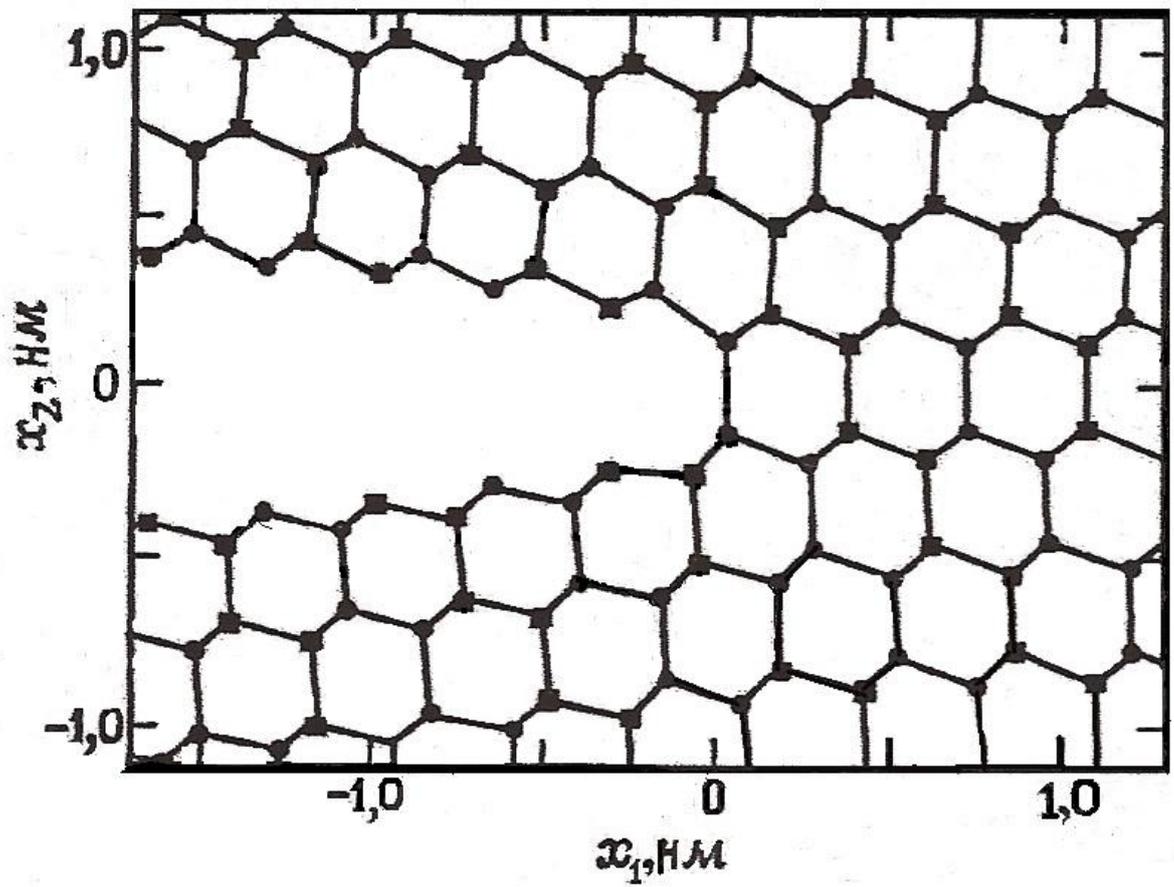

Fig.2. Computer model of crack tip structure for two parallel planes (011), body-centered cubic iron lattice. [3]

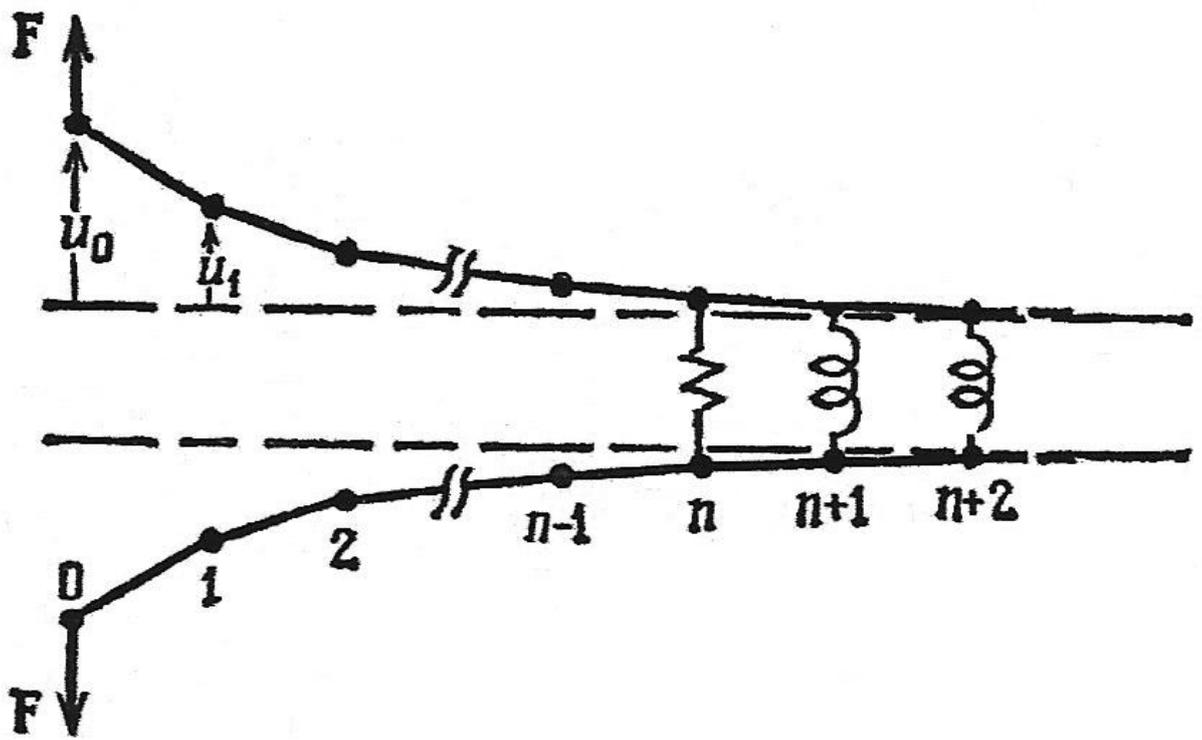

Fig. 3. One-dimensional crack model [4]

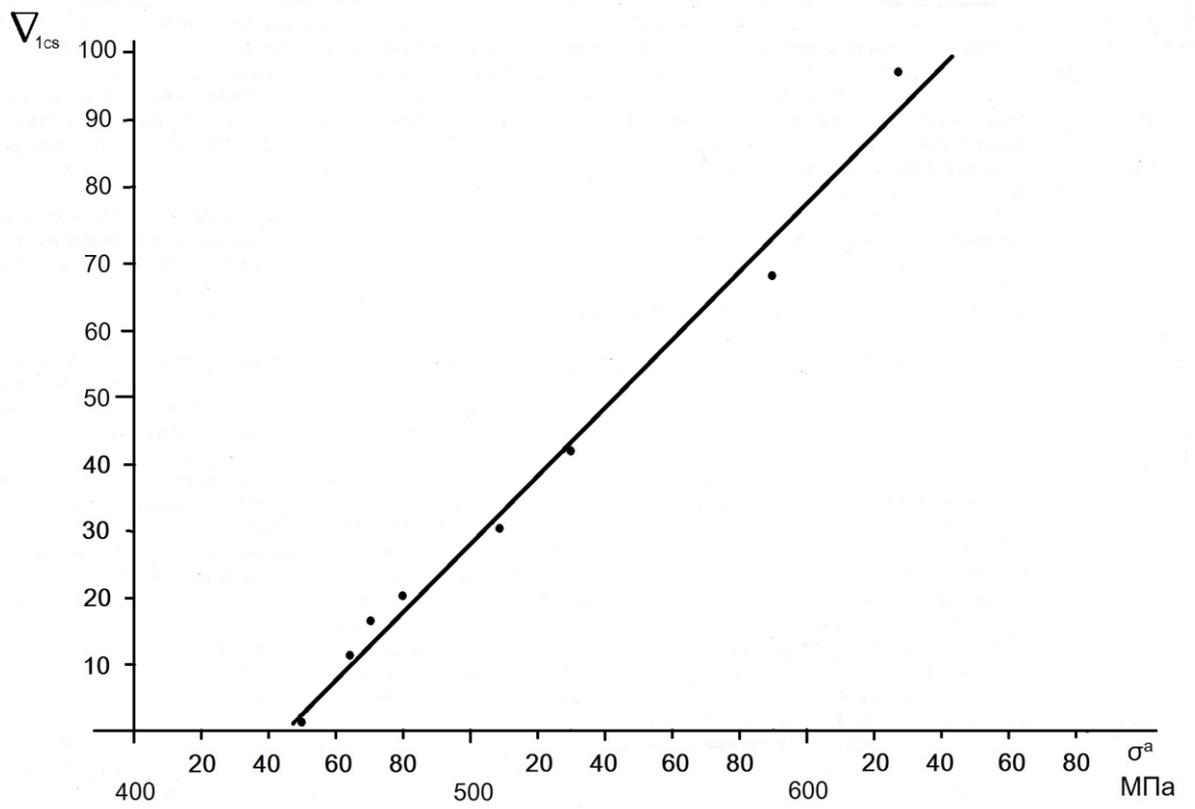

Fig. 4. Dependence $\nabla_{1cs}$ - average number of interplanar bonds being destroyed within 1 cycle of $\sigma^a$ (peak stress).

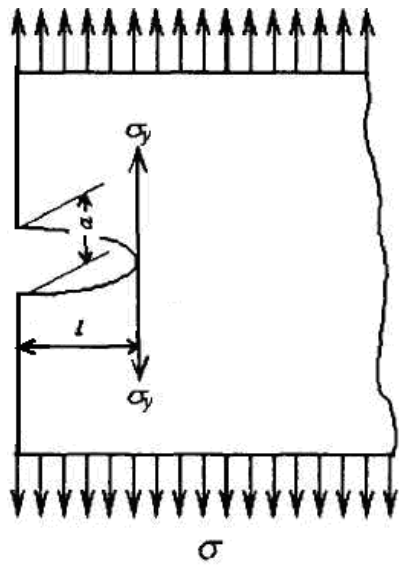

Fig. P1. A plate with a side crack having opening a and length l under stress σ.

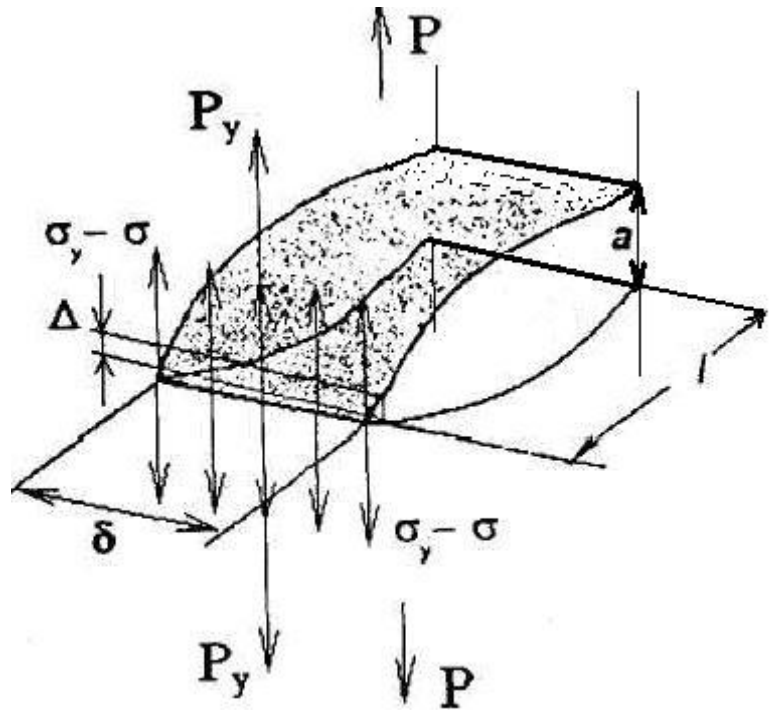

Fig. P2. A side crack model in the plate presented in Fig.P1